\documentclass[12pt,preprint]{aastex}
\usepackage{graphics,graphicx}
\begin{document}
\title{Non-WKB Models of the FIP Effect: The Role of Slow Mode Waves}


\author{J. Martin Laming}
\affil{Space Science Division, Naval Research Laboratory Code 7674L,
Washington, D.C. 20375} %\email{laming@nrl.navy.mil}

\begin{abstract}
A model for element abundance fractionation between the solar chromosphere and
corona is further developed. The ponderomotive force due to Alfv\'en waves
propagating through, or reflecting from the chromosphere in solar conditions generally
accelerates chromospheric ions, but not neutrals, into the corona. This gives rise
to what has become known as the First Ionization Potential (FIP) Effect.
We incorporate new physical processes into the model. The chromospheric ionization
balance is improved, and the effect of different approximations is discussed. We also
treat the parametric generation of slow mode
waves by the parallel propagating Alfv\'en waves. This is also an effect of the
ponderomotive force, arising from the periodic variation of the magnetic pressure driving
an acoustic mode, which adds to the background longitudinal pressure. This can have subtle
effects on the fractionation, rendering it quasi-mass independent in the lower regions of the
chromosphere. We also briefly discuss the change in the fractionation with Alfv\'en
wave frequency, relative to the frequency of the overlying coronal loop resonance.
\end{abstract}


\section{Introduction}
The First
Ionization Potential (FIP) effect is the by now well known enhancement
in abundance by a factor of 3-4 over photospheric values of elements
in the solar corona with FIP less than about 10 eV. These low FIP elements
include Fe, Si, Mg, etc. Elements with FIP greater than 10 eV
(high-FIP) mostly retain their photospheric composition. This
was actually first observed in the
1960's (Pottasch 1963), making it nearly as old as the problem of
coronal heating. It has been taken seriously as a phenomenon since the
mid 1980's (Meyer 1985ab). There are a number of studies of the FIP effect in
different regions of the solar corona and wind, reviewed
in Feldman \& Laming (2000) and Laming (2004), and references
therein. With the launch of the Extreme Ultraviolet Explorer (EUVE) in the
1990's, it became clear that element abundances in late-type stellar coronae
also do not always resemble the stellar photospheric composition. The FIP
effect is also observed in many solar-like late type stars. At higher
activity levels and/or later spectral types, a so called ``inverse
FIP'' effect is observed, where the low FIP elements are depleted in
the corona \citep[e.g.][]{wood10}.

A variety of models have been proposed to explain these phenomena.
\citet{laming04,laming09} review many of these different scenarios, and argue that
the ponderomotive force is the most likely agent of FIP fractionation. This
force arises as Alfv\'en waves propagate through the chromosphere, and acts on
chromospheric ions, but not neutrals. Physically, it corresponds to the interaction of waves and plasma through the refractive index of the medium. Waves carrying significant energy and momentum refracting or reflecting in a plasma must exert a force, in this case on the charged particles that contribute to the dielectric
tensor, but not on the neutrals. The ponderomotive force in the chromosphere may
in principle be directed upwards or downwards, giving rise to FIP or so-called ``inverse FIP''
effects respectively (i.e. a coronal enhancement or depletion of low FIP ions).

The chromosphere-corona interface is generally a barrier to Alfv\'en wave
propagation; upcoming waves from the chromosphere are usually reflected back down again, and downward directed waves from the corona reflect back upwards, as
illustrated in the right hand footpoint (chromosphere B) in Figure 1.
Alfv\'en waves with predominantly coronal origin generally
give rise to the positive (i.e. solar-like) FIP effect, while 
waves generated by upward propagating acoustic waves associated with
stellar convection may produce inverse FIP effect.
%Since
%upward propagating acoustic waves principally generate Alfv\'en waves beginning
%at the layer where the sound speed and Alfv\'en speed are equal
%\citep[plasma beta $\beta =8\pi P/B^2=1.2$, where $P$ is the gas pressure
%and $B$ is the magnetic field, e.g.][]{cally08}, the observed positive FIP effect in
%stars of similar spectral type to the Sun suggests that upward
%propagating acoustic waves in these stars do not reach this layer
%before being reflected downwards again, while in later type stars with
%stronger magnetic fields giving $\beta =1.2$ at lower altitudes, they
%do.
The association of coronal abundance anomalies with Alfv\'en waves
gives us a unique and unexpected diagnostic with which to explore the
behavior of MHD turbulence in solar and stellar upper atmospheres. While we will
argue that the coronal Alfv\'en waves themselves are actually byproducts of
processes that heat solar and stellar coronae, (most likely accomplished by
various forms of ``nanoflares''), an understanding of coronal
abundance anomalies still becomes far more central to exploring
coronal heating than would be the case in prior models for the
fractionation invoking thermal processes such as diffusion.

In this paper we seek to develop the ponderomotive force model incorporating
the parametric generation of parallel propagating slow mode waves by the
Alfv\'en waves themselves, together with revisions to some of the atomic data.
First, in section 2, we review the important features of the \citet{laming04,laming09}
model. Section 3 outlines the various theoretical
refinements made in this paper, and section 4 describes new results for
fractionations in open and closed magnetic field configurations. Section 5 discusses in
more detail the effect of the parametric generation of slow mode waves
and its implications for
fractionation in closed coronal loops and in the slow speed solar wind. We also
consider limits
on the upward flow speed through the chromosphere, and make final conclusions.

\section{The Ponderomotive Model Revisited}
The ponderomotive force stems from second order terms in $\delta {\bf
J}\times\delta {\bf B}/c$ and $\rho\delta {\bf v}\cdot\nabla\delta{\bf v}$
in the MHD momentum equation. It can be
manipulated (e.g Litwin \& Rosner 1998, Laming 2009) for waves of
frequency $\omega _A<< \Omega$, the ion gyrofrequency, into the time
averaged form
\begin{equation}
F={\partial\over\partial z}\left(q^2\delta E_{\perp}^2\over 4m\Omega ^2\right)
={mc^2\over 4}{\partial\over\partial z}\left(\delta E_{\perp}^2\over B^2\right),
\end{equation}
where $\delta E_{\perp}$ is the perpendicular wave electric field and $q$ is
the ion charge. The ponderomotive acceleration, $F/m$, is independent of
the ion mass. The Laming (2004, 2009) model comes about as a natural
extension of existing work on Alfv\'en wave propagation in the solar
atmosphere with essentially no extra physics required.  It is also the
model most worked out in detail to interpret observations, giving it
unique potential for diagnosing wave processes in the corona and
chromosphere.

The basic model builds on Hollweg (1984), where upward propagating
Alfv\'en waves were introduced at one loop footpoint. Here they could
reflect back down into the chromosphere, or be transmitted into the
loop, where they propagated back and forth with a small probability of
leaking back into the chromosphere at each end. With reference to Figure 1,
we initiate our simulations with one downward propagating wave at the
$\beta\sim 1.2$ layer in chromosphere A, and integrate the Alfv\'en wave
transport equations (see below) to chromosphere B to evaluate the standing wave pattern there.
The chromosphere at each footpoint can be based on any of the \citet{vernazza81}
models or similar. Here we use the \citet{avrett08} update of
VALC. The ionization balance of the minor ions is computed at each
height in the chromosphere using the model temperature and electron
density, and a coronal UV-X-ray spectrum appropriately absorbed in
the intervening chromospheric layers. We have tried a number of
different spectra based on \citet{vernazza78}, or model flare spectra
computed using CHIANTI \citep[see e.g.][for the 2000 Bastille Day flare]{huba05}.
The atomic data are
as in \citet{laming04} and \citet{laming09}, with estimates for the
charge exchange ionization for Si, Fe, and other low FIP elements
added (see subsection 3.3). The chromospheric magnetic field is taken to
be a 2D force free field from Athay (1981) and designed to match chromospheric
magnetic fields in Gary (2001) and Campos \& Mendes (1995), which represents the
expansion of the field from the high $\beta$ photosphere where the field is
concentrated into small network segments, into the low $\beta$ chromosphere where the
field expands to fill much more of the volume.

We model the Alfv\'en waves in a non-WKB approximation. The procedure
follows that described in detail by Cranmer \& van Ballegooijen
(2005), but applied to closed rather than open magnetic field
structures. The curvature of the loop is neglected. For Alfv\'en waves, where the
energy flux is necessarily directed along the field direction, this is unlikely to have
a significant effect. Some extra damping may result as the wave develops a component of its
wavevector perpendicular to the field, but we neglect this and all other damping mechanisms
in this work. The transport equations are
\begin{equation} {\partial z_{\pm}\over\partial t}+\left(u\pm
V_A\right){\partial z_{\pm}\over\partial r}= \left(u\pm
V_A\right)\left({z_{\pm}\over 4H_D}+{z_{\mp}\over 2H_A}\right),
\end{equation}
where $z_{\pm}=\delta v_{\perp}\pm \delta B_{\perp}/\sqrt{4\pi\rho }$ are the
Els\"asser variables for Alfv\'en waves propagating against or along
the magnetic field respectively, and are valid for torsional or planar Alfv\'en waves.
The Alfv\'en speed is $V_A$, the
upward flow speed is $u$ and the density is $\rho $. The signed scale
heights are $H_D=\rho /\left(\partial\rho /\partial r\right)$ and
$H_A=V_A/\left(\partial V_A/\partial r\right)$. In the solar
chromosphere and corona $u << V_A$, and we put $u=0$. The calculation of $V_A$ uses the total
(ionized and neutral) gas density, since the wave frequencies of interest here are well
below the charge exchange rate, and neutrals are well coupled to the wave. Charge changing
collisions involving electrons (impact ionization, and radiative and dielectronic recombination)
are generally slower than charge exchange in chromospheric conditions. Hence in considering
the wave propagation, ion-neutral friction is neglected, though it is included in the
evaluation of the fractionation.

The fractionations are calculated by
postprocessing the non-WKB wave. This is
valid because the fractionation has a negligible effect on the wave
propagation. The degree of fractionation is given by the formula
\begin{equation}
{\rm fractionation} =\exp\left(2\int
_{z_l}^{z_u}{\xi _sa\nu _{eff}/\nu _{s,i}/v_s^2}dz\right)
\end{equation}
(see
equation 9, Laming 2009, equation 12, Laming 2004, which follow
Schwadron et al. 1999), where $\xi _s$ is the ionization fraction of
element $s$, $a$ is the ponderomotive acceleration, $\nu _{eff}=\nu
_{s,i}\nu _{s,n}/\left[\xi _s\nu _{s,n}+\left(1-\xi _s\right)\nu
_{s,i}\right]$ where $\nu _{s,i}$ and $\nu _{s,n}$ are the collision
rates of ions and neutrals, respectively, of element $s$ with the
ambient gas. Also $v_s^2 =kT/m_s +v_{\mu turb}^2 +v_{turb}^2$, where
$v_{\mu turb}$ is the amplitude of microturbulence coming from the
\citet{avrett08} chromospheric model, and $v_{turb}$, discussed
further in section 3.2, is the amplitude of longitudinal waves induced
by the Alfv\'en waves themselves. The limits of integration, $z_l$ and $z_u$ are
the lower and upper limits over which the ponderomotive force acts. We take
$z_l$ to be where $\beta \simeq 1$, and $z_u$ is in the transition region at an
altitude where all elements are ionized.

\section{New Model Features}
\subsection{Slow Mode Waves}
We have introduced an extra longitudinal pressure associated with the Alfv\'en waves
proportional to $v_{turb}^2$, which has the effect of causing
some saturation of the FIP fractionation. Here we give the physical motivation for this
extra term,
which arises from the generation of slow mode waves. Physically, the periodic
variation in magnetic pressure of the Alfv\'en wave drives longitudinal compressional
waves. These generated acoustic waves can act back on the Alfv\'en driver, as the
compressional wave introduces a periodic variation in the Alfv\'en speed, which
generates new Alfv\'en waves. We illustrate the generation of slow mode or acoustic
waves by the ponderomotive force associated with plane Alfv\'en waves with a simple 1D
calculation. The linearized momentum equation keeping
terms to all orders in perturbed quantities is (all symbols have their usual meanings),
\begin{equation}
\left(\rho +\delta\rho\right){\partial\delta v_z\over\partial t}+
\left(\rho +\delta\rho\right)\delta v_z{\partial\delta v_z\over\partial z}=
\left(\rho +\delta\rho\right){\partial\over\partial z}{\delta B^2\over 8\pi\left(\rho +\delta\rho\right)}-{\partial\delta P\over\partial z} -g\delta\rho ,
\end{equation}
where
\begin{eqnarray}
\delta\rho &=-\rho\nabla\cdot{\bf \xi}-\xi _z{\partial\rho\over\partial z} =-\rho\xi\left(ik_s+{1\over L_{\rho}}\right)\cr
\delta P &=-\gamma P\nabla\cdot{\bf \xi}-\xi _z{\partial P\over\partial z}
=-P\xi\left(ik_s\gamma +{1\over L_P}\right)
\end{eqnarray}
for Eulerian displacement ${\bf \xi}\propto\exp i\left(\omega  _st+k_sz\right)$,
where $L_p=P/\left(\partial P/\partial z\right)$ and
$L_{\rho}=\rho /\left(\partial\rho /\partial z\right)$
(signed) pressure and density scale heights respectively. The first term on the
right hand side of equation 4 represents the instantaneous ponderomotive force. In cases
where the WKB approximation applies, $\delta B\propto\rho ^{1/4}$, and this expression is
equivalent to the more usual form $-\partial\left(\delta B^2/8\pi\right)/\partial z$.
Substituting for $\delta\rho$ and $\delta P$ from equations 5,
keeping terms as high as second order in small quantities, equation 3 becomes
\begin{equation}
-i{\omega _s\over L_{\rho}}\delta v_z^2+\left(-\omega _s^2 +k_s^2c_s^2-i{k_sc_s^2\over L_P}-
{c_s^2\over\gamma L_P^2}-i{k_sc_s^2\over\gamma L_P}-ik_sg-{g\over L_{\rho}}\right)\delta v_z
-i\omega _s{\partial\over\partial z}{\delta B^2\over 8\pi\rho}=0.
\end{equation}
This is considerably simplified in isothermal conditions, $\gamma =1$, $L_P=L_{\rho}=-c_s^2/g$, so that
\begin{equation}
-i{\omega _s\over L_{\rho}}\delta v_z^2+\left(-\omega _s^2 +k_s^2c_s^2+ik_sg\right)\delta v_z
-i\omega _s{\partial\over\partial z}{\delta B^2\over 8\pi\rho}=0.
\end{equation}
We put $\Im k_s=-g/2c_s^2$, and $\sqrt{\left(\Re k_s\right)^2+g^2/4c_s^4}
=2\Re k_A/n$, $\omega =2\omega _A/n$, where $k_A$ and $\omega _A$ are the wavevector and angular frequency of the Alfv\'en wave with $n=1,2,3,..$
(anticipating the result below). We find
\begin{equation}
\delta v_z^2-\delta v_z iL_{\rho}\omega _s\left(1-{c_s^2\over V_A^2}\right)+
L_{\rho}{\partial\over\partial z}{\delta B^2\over 8\pi\rho}=0
\end{equation}
with solution
\begin{equation}
\delta v_z={-i\over 2}\left[\sqrt{\delta v_A^2+L_{\rho}^2\omega _s^2
\left(1-{c_s^2\over V_A^2}\right)^2}-L_{\rho}\omega _s\left(1-{c_s^2\over V_A^2}\right)\right]
\end{equation}
where we have put ${\partial\over\partial z}\left(\delta B^2/ 8\pi\rho\right)=
\left(\delta B^2/4\pi\rho\right)/4L_{\rho}=\delta v_A^2/4L_{\rho}$ using the
WKB result for Alfv\'en waves in a density gradient, and assuming $1/L_{\rho} >> \Re k_A$.
When $c_s^2\sim V_A^2$ or $L_{\rho}\rightarrow 0$, $\delta v_z\simeq-i\delta v_A/2$. In
the opposite limit $\delta v_z\simeq -i\delta v_A^2/4L_{\rho}\omega _s
\left(1-c_s^2/V_A^2\right)$. In these two cases $\omega _s=
\omega _A$ or $\omega _s=2\omega _A$ respectively. In fact acoustic waves can be
generated with $\omega _s=2\omega _A/n$, with higher $n$ becoming more intense as the
nonlinearity increases \citep{landauCM}. \citet{vasheghani11} treat the case of slow
mode wave generation by a torsional Alfv\'en wave. This is subtly different to the case of
a plane Alfv\'en wave considered here, and the FIP fractionation resulting from such a model
will be investigated in a future paper.

Anticipating applications to possibly nonlinear Alfv\'en and slow mode wave
amplitudes, we revisit the analysis above retaining more higher order terms.
From $\delta\rho =-\rho\xi\left(ik_s+1/L\right)$  we derive
\begin{equation}{\partial\delta\rho\over\partial z}=\delta\rho\left(ik_s+
{1\over L}\right).
\end{equation}
which when substituted into the linearized continuity equation,
\begin{equation}
{\partial\delta\rho\over\partial t}+{\partial\over\partial z}\left(\rho
\delta v_z+\delta\rho\delta v_z\right)=0,
\end{equation}
with the time derivatives $\partial\delta v_z/\partial t=i\omega_s\delta v_z$ and
$\partial\delta\rho /\partial t=i\omega_s\delta\rho $,
gives
\begin{equation}
\delta\rho =-\left(ik_s+1/L_{\rho}\right){\rho\delta v_z\over i\omega _s}
\left(1+{2k_s\delta v_z\over\omega}+{\delta v_z\over i\omega L_{\rho}}\right)^{-1}.
\end{equation}
Writing $\delta P=\gamma P\left(\delta\rho /\rho+\delta v_z/i\omega L_{\rho}\right)
-P\delta v_z/i\omega L_P$ we similarly derive
\begin{equation}
{\partial\delta P\over\partial z}=\left({1\over L_P}+ik_s\right)
\left(c_s^2\delta\rho +P{\delta v_z\over i\omega}\left({\gamma\over L_{\rho}}-{1\over L_P}\right)\right).
\end{equation}
We now eliminate $\delta \rho$ and $\delta P$ in favor of $\delta v_z$
in equation 3 to derive a quartic equation in $\delta v_z$ for the driven slow mode wave with angular frequency $\omega _s$ and wavevector $k_s$;
\begin{eqnarray}
&\delta v_z^4\left[-{k_s^3\over\omega _s}\right]+\delta v_z^3\left[-3k_s^2+\left({c_s^2\over L_{\rho}}-{c_s^2\over\gamma L_p}\right)\left({1\over L_p}+ik_s\right)\left({2k_s^2\over\omega _s^2}
+{k_s\over i\omega _s^2L_{\rho}}\right)\right] \cr
+&\delta v_z^2\left[
-3k_s\omega _s +\left({c_s^2\over L_{\rho}}-{c_s^2\over\gamma L_p}\right)\left({1\over L_p}+ik_s\right)\left({2k_s\over\omega _s}
+{1\over i\omega _sL_{\rho}}\right)+{k_s^3c_s^2\over\omega _s}-
i{k_s^2c_s^2\over\omega _sL_p}-{k_sc_s^2\over\gamma\omega L_p^2}\right]\cr
+&\delta v_z^2\left[-i{k_s^2c_s^2\over\gamma\omega _sL_p}-i{k_s^2g\over\omega _s}-{gk_s\over\omega _sL_{\rho}}+\left(2ik_s+{1\over L_{\rho}}\right)
\left(ik_A\delta v_A^2{k_s\over\omega _s}+{\delta v_A^2k_s\over 2\omega L_{\rho}}+{\delta v_A^2ik_s^2\over 2\omega _s}\right)\right]\cr
+&\delta v_z\left[-\omega _s^2+k_s^2c_s^2-i{k_sc_s^2\over L_p}-{c_s^2\over\gamma L_p^2}-ik_s{c_s^2\over\gamma L_p}-ik_sg-{g\over L_{\rho}}+\left(3ik_s+{1\over L_{\rho}}\right)ik_A\delta v_A^2+{i\delta v_A^2k_s\over 2L_{\rho}}-{\delta v_A^2k_s^2\over 2}\right]\cr
& -\omega _sk_A\delta v_A^2=0.
\end{eqnarray}
To lowest order, the terms in $\delta v_z$ and the constant are the same as in equation 6.
The quadratic and higher terms are changed, because of the difference between equations 12
and 13, and those in equation 5.
Inserting even  accurate spatial derivatives in place of those in equations 12 and 13
would generate yet more higher order terms, extending $n$ in principle without limit.
\citet{landauCM} give a similar conclusion in their treatment of parametric resonance.

We solve equation 14 numerically, with the same $k_s$ (real and imaginary
parts) as above. We select the solution with the lowest absolute magnitude as the
physically correct solution for our problem, this being the solution that goes to zero
as $\delta v_A \rightarrow 0$. This is always close to the solution obtained discarding all
terms of order higher than $\delta v_z^2$ in equation 14, and usually close to the
case when the term in $\delta v_z^2$ is also neglected. This describes
turbulence for parallel propagating waves. \citet{zaqarashvili06} give a detailed treatment
of the interaction between weak Alfv\'en and sound waves in a homogeneous medium, where
acoustic and Alfv\'en speeds are equal. The stronger generation of acoustic waves by the
ponderomotive force in a density gradient is demonstrated by the simulations of
\citet{delzanna05}, where slow mode waves are seen propagating up from loop footpoints
with properties consistent with the solutions of equations 9 or 14.

Closer to the layer where $c_s^2=V_A^2$,
the magnetic field becomes more curved, giving rise to higher
perpendicular components of Alfv\'en wave wavevectors, and potentially
stronger turbulent cascade and/or mode conversion. In this case we expect stronger slow mode waves.
In \citet{laming09} we assumed $\delta v_s=\delta v_A$. However considerations of mode
conversion for initially upward propagating acoustic waves suggest that higher slow
mode intensities than this should be present. \citet{khomenko11} report that at conditions for
maximum conversion of a high $\beta$ fast mode wave to a low $\beta$ Alfv\'en wave, the
low $\beta$ slow mode wave has 2-3 times more flux. In this paper, we make the approximation
\begin{equation}
\delta v_s^2 = \delta v_z^2 +6\delta v_A^2c_s^2/V_A^2,
\end{equation}
where $\delta v_z$ represents the solution of equation 14, and the factor 6 in the last term is
motivated by calculations illustrated in \citet{cranmer07} and \citet{khomenko10}. Slow mode waves
governed by equation 15 have the effect of suppressing fractionation in the low chromosphere
close to where the plasma $\beta\simeq 1.2$. Studies of mode conversion between acoustic and Alfv\'en
waves generally show the upward moving acoustic wave beginning to convert to Alfv\'en waves at
the $\beta\simeq 1.2$ layer, and mode conversion continuing over a range of altitudes of order
100's of km \citep[e.g][]{cally08}. The explicit incorporation of such effects is well beyond the
scope of the work here, and the prescription in equation 15 should be sufficient to avoid
the occurrence of unphysical fractionations. Even so, it remains a feature of this work requiring
further investigation in future papers.

\subsection{Normalization Relative to Oxygen}
In previous papers \citet{laming04,laming09} we have discussed FIP fractionations
relative to H. However the derivation of 3 has assumed fractionated elements are
minor species, with the fractionation having no back reaction on the flow due to the neglect of inertial terms. Thus it is more appropriate to present and describe
element fractionations with respect to another minor element. We choose O, which is a common choice for observers also.

\subsection{Charge Exchange Ionization}
Charge exchange rates have been previously taken from the compilation
of \citet{kingdon96}. These have been supplemented more recently with
charge exchange ionization rates for Si and Fe, colliding with protons.
We implement an estimate of the charge exchange ionization rate for all ions
with lower FIP than H as follows. We estimate the radius $R$ at which the sum of the
binding energy of the electron in the target neutral and the polarization potential energy
of the target neutral in the electric field of the incoming proton and equal to the
equivalent sum for the resulting neutral H atom in the electric field of the newly formed ion,
(known as the radius of the potential crossing) from
\begin{equation}
V=-{\left(\alpha _s-\alpha _H\right) \over R^4}=-I_H +I_s
\end{equation}
where $V$ represents the difference in potential energy of the proton in terms of the
polarizability $\alpha _s$ of the target atom and resulting ion in terms of $\alpha _H$,
the polarizability of the resulting
hydrogen atom. $I_H$ and $I_s$ are the ionization potentials of
hydrogen and the target atom, $s$, respectively. Polarizabilities and ionization
potential here are in atomic units. The cross section is then $\sigma _{cxi}=
\pi R^2/2=\pi\sqrt{\left(\alpha _s-\alpha _H\right)/\left(I_H-I_s\right)}/2$,
assuming the maximum probability for a reaction is 1/2. This estimate is a slightly
different form of the Langevin formula given by
\citet{ferland97}. This approximation gives good agreement with the calculations
of \citet{allan88} for charge exchange ionization of Mg. These authors comments
that similar rates should exist for all other elements with ionization potentials
below that of H, and we apply it to all of these elements. The charge exchange implemented is
just the thermal process. No account is taken of any possible effects of the waves on the
chromospheric ionization balance.

\subsection{Comparison of Different Approximations}
\citet{carlsson02} have argued that the concept of an ``average''
chromosphere pursued by \citet{avrett08} and its antecedents is
invalid, due to the extreme dynamics associated with chromospheric
shock waves, arguing that ``the mean value of a dynamic property is not the same as
that property evaluated for the mean atmosphere''.
\citet{avrett08} derive mean values for plasma properties based on
observations, not on calculated mean atmospheres. More importantly,
\citet{carlsson02} show that the ionization fraction for H varies very
little about its mean, due to the length of ionization and
recombination times compared to the frequencies of shocks, and that
their average electron density agrees very well with that in the
\citet{vernazza81} model C. It is easy to see that other high FIP elements
should behave similarly, and that the ionization balance we calculate (the
overwhelmingly most important chromospheric property to us) should
not be greatly in error, if at all. The inclusion of charge exchange ionization
(section 3.3) increases the ionization level for all other elements as the ionization
of hydrogen is increased above its thermal equilibrium level.
The Ca$^+$ to Ca$^{2+}$ ionization balance is considered by \citet{wedemeyer11}.
This is more variable than that for H to H$^+$ in \citet{carlsson02}, but is less of
a concern to us, since the ponderomotive force experienced by an ion is independent of
ion charge, so long as
$\omega _A<< \Omega$.

We give sample calculations that go some way to quantifying
the effect of these issues. Figure 2 shows the coronal section of the 100,000 km
long loop with magnetic field $B=20$G. The density at the loop apex is $5\times 10^8$ cm$^{-3}$.
This gives a resonant angular
frequency of 0.07 rad s$^{-1}$. A wave of this frequency propagates on the loop, which
is thus half a wavelength long. The top panel gives
the Els\"asser variables (real and imaginary parts), the middle panel gives the wave
energy fluxes and their difference, and the third panel gives the ponderomotive
acceleration. Figure 3 shows the chromospheric section of the loop
where the fractionations are evaluated. This has the same three panels as for
Figure 2, where the third panel also gives the slow mode wave amplitude, with a fourth panel
(bottom right) that gives FIP fractionation and the
ionization fraction of elements Fe, Mg, S, and He relative to O. In Table 1 fractionations for He, C, N, Ne, Mg, Si, S, Ar, and Fe relative to O are displayed, calculated according to our
basic model described above, labeled ``baseline''. In the succeeding columns, we
give FIP fractionations calculated with different modifications to the model.
In the first case the density given by \citet{avrett08} is
modified so that the model density is consistent with the H ionization fraction and
the assumption of photoionization-recombination equilibrium. The second variation
gives fractionations calculated assuming the ionization fractions to be given by
the Saha equation at the temperature and density in the chromospheric model. The first
variation reduces the degree of ionization in the chromosphere, while the second one
increases it. This has the opposite effect on the fractionations, since these are
given relative to O, and the increase or decrease of the ionization of O has a
bigger effect on its fractionation than is the case with the other elements.
As can be seen, the basic phenomenon of the FIP effect remains unaffected by
the choice of approximation,
but some of the details, e.g. the Mg/Fe ratio are subtly different. We return to
discuss this in more detail below, in subsection 4.2.

The final column in Table 1 gives the FIP fractionation calculated with the
assumption of slow mode wave amplitude $\delta v_z = \delta v_A$, the amplitude of
the Alfv\'en wave, as taken in \citet{laming09}, instead of implementing the solution
of equation 14. The overall fractionation is reduced
for the same Alfv\'en wave amplitude, due to the increase in longitudinal pressure
with the higher slow mode wave amplitude. In \citet{laming09},we found that the
FIP Effect saturated in this case at values around 3-4 for arbitrarily high Alfv\'en
wave amplitude. In section 4, we will consider the behavior of the FIP Effect with
resonant and nonresonant waves.

\section{Results}
\subsection{Coronal Hole}
A coronal hole is modeled int he same fashion. The chromosphere is identical to the case above,
and the density evolves smoothly off-limb, declining to about $2\times 10^6$ cm$^{-3}$ at
an altitude of $5\times 10^5$ km \citep[as in e.g. Figure 5 of][]{cranmer09}. The magnetic
field follows the model of \citet{banaskiewicz98}.
We take the coronal hole Alfv\'en wave spectrum calculated by \citet{cranmer07} as our
starting point. It is illustrated in their Figure 3 for a position in the solar
transition region. It is represented by the five wave frequencies and amplitudes at
the starting position of the non-WKB integration of the wave transport equations, in this
case at an altitude of 500,000 km. Parameters are chosen to match Figure 9 in
\citet{cranmer07}, and are
given in Table 2 as the spectrum labeled ``v0''. Figure 4 shows the wave amplitudes in the
coronal hole section of the calculation, with the same three panels as for Figure 2, but
illustrated up to an altitude of 500,000 km. Figure 5 shows the chromospheric response with
the same four panels as in Figure 3. The ponderomotive acceleration (Figure 5, top right)
has a similar ``spike'' at the top of the chromosphere as in Figure 3 for the closed loop,
but is about an order of magnitude weaker. Lower down, the
Alfv\'en wave amplitudes, and the corresponding slow mode wave amplitudes are larger than
before. The net effect of smaller ponderomotive acceleration and higher slow mode
wave amplitude is to reduce the FIP effect from the case in Figure 3. The resulting
fractionation here is very similar to that found by \citet{cranmer07}.
In models v1-2 in Table 2, we increase the amplitude of the highest and lowest frequency wave
in the spectrum respectively. Increasing the highest frequency wave amplitude has
the effect of increasing the fractionation predicted, while increasing the low frequency wave has
very little effect. The
results from all three models are summarized in Table 3, and compared with observations
for various open field regions. Reasonable agreement for the fractionation of low FIP
elements is seen. A small depletion of He is seen, but not as large as observed.
Our purpose here has not been to provide a definitive calculation of the FIP effect in a
coronal hole, but merely to illustrate that the observed difference between the FIP effect
in closed and open field lines arises naturally in this model.

\subsection{Closed Loop}
We now explore fractionation in a closed coronal loop, illustrating the difference that
a coronal resonant frequency can make to the fractionation. In Table 4 we give the
fractionations computed for a closed loop with length 100,000 km, and a magnetic field
of 20 G. A coronal wave at the loop resonant angular frequency of 0.07 rad s$^{-1}$ is
modeled. The wave transport equations (equation 2) are integrated from the $\beta =1$ layer
in one chromosphere, where and initially downgoing wave amplitude is specified, through the loop
to the opposite chromospheric footpoint, where the FIP fractionations are evaluated. The
initial wave amplitudes are
0.4, 0.5, and 0.6 km s$^{-1}$, giving coronal wave peak amplitudes of approximately 55, 70,
and 82 km s$^{-1}$ respectively.

These cases illustrate the variation of the FIP effect with wave amplitude.
The six columns on the right hand side give various observations of FIP fractionation.
\citet{zurbuchen02} and \citet{vonsteiger00} both give fractionations measured
{\it in situ} in the
solar wind over relatively long periods of time. \citet{giammanco07,giammanco08} give
fractionations also measured in the solar wind, but over time periods selected such that
the wind speed was close to 380, 390 or 400 km s$^{-1}$.
\citet{phillips03,sylwester10a,sylwester10b,mckenzie94} give abundance ratios
measured spectroscopically in solar flares, and \citet{bryans09} observe quiet solar
corona above the western limb, also spectroscopically. While the agreement between the
FIP fractionations calculated for the
initial Alfv\'en wave amplitude corresponding to 0.5 km s$^{-1}$ is generally good, there
are some important discrepancies. The ratio C/O is typically observed  in the solar wind
to be higher than calculated, as is S/O for some of the observations. While Fe/O and Mg/O
are reasonably well reproduced, the direct
ratio between Fe and Mg is not, except in the case of \citet{bryans09}. The last case,
with initial wave amplitude 0.6 km s$^{-1}$, is designed to match these observations,
and does quite well. Only K is seriously discrepant, with S also somewhat underestimated.
In the solar wind and flare observations, Fe and Mg are often fractionated by the same
amount.

Table 5 gives FIP fractionations for varying Alfv\'en wave frequency, with the
amplitudes chosen to keep the Fe/O abundance ratio close to 4 as is often observed. Here, we
have also included a wave field of chromospheric origin designed to have the same
properties as that in the v0 model for the open field case. Downgoing wave amplitudes are specified
at the $\beta =1$ layer in one chromosphere, and then the wave transport equations are
integrated back to the opposite footpoint where FIP fractionations are evaluated. The
column corresponding to the resonant frequency of 0.07 rad s$^{-1}$ is the same as in
Table 4, but for the new model. The chromospheric waves can be seen to have rather little
effect, with the biggest changes being seen in the increased fractionations of C/O and S/O.
Moving away from this resonance, either to lower or higher frequency, the
C/O and S/O ratios increase to better agree with observations. At higher frequency so too does
the Mg/O ratio, so that Mg and Fe fractionate more closely to the same degree; quasi-mass independent
fractionation is achieved. Going to yet higher frequency,
all high FIP elements remain unchanged, while all low FIP elements are enhanced by a
factor 3-4. The depletions of He and Ne are lost.
The quiet sun observations of \citet{bryans09} are best matched at the frequency of 0.075
rad s$^{-1}$
and with the exception of K, the consistency between their measurements and
the model is excellent at an initial wave amplitude of 0.7 km s$^{-1}$.

Figures 6 and 7 illustrate the ponderomotive acceleration and FIP fractionation within the
chromosphere for the cases of frequencies 0.06, 0.07, 0.085 and 0.105 rad s$^{-1}$. At
0.06 rad s$^{-1}$, the ponderomotive acceleration is positive from about 800 km up,
and has a ``spike'' at about 2150 km, with maximum close to $10^6$ cm s$^{-2}$. Fractionation of
Fe, Mg, and S is similar low down, but the fractionation hierarchy Fe $>$ Mg $>$ S is established
in the range 1500-2000 km. For 0.07 rad s$^{-1}$, the ponderomotive
acceleration has a similar,
but slightly larger maximum at about 2150 km. In response to this Fe, Mg and S undergo a similar
and small inverse FIP fractionation up to 1500 km, giving way to positive FIP
higher up. The fractionation pattern Fe $>$ Mg $>$ S is even stronger here than for
0.06 rad s$^{-1}$, and is mainly occurring at the ``spike'' in the ponderomotive acceleration.
In the last two cases, the ponderomotive force is stronger lower down in the chromospheric,
and the ``spike'' at 2150 km becomes less pronounced. The slow mode wave amplitude is
also stronger lower down. At 0.105 rad s$^{-1}$, all
fractionation occurs by 1600 km, and the local maximum in the ponderomotive acceleration
at 2150 km has no effect.

These fractionation patterns have simple qualitative explanations. First we display in Figure 8,
the 0.07 rad s$^{-1}$ case again to illustrate the relation of the fractionation
to important features of the chromosphere. The left panels give the ponderomotive acceleration
(bottom) and the density and temperature structure of the chromosphere (top). The ``spike''
in the ponderomotive acceleration can be seen to stem from the steep density gradient beginning
at an altitude of about 2100 km. The solid and dashed lines in the bottom plot show the ponderomotive
acceleration with and without the energy loss to slow mode waves. In the regions where significant
fractionation occurs, the slow mode wave do not affect the ponderomotive acceleration very much,
and their main effect on the fractionation is through the additional longitudinal pressure they
provide. The panels on the right hand side show the same fractionations as before (bottom),
and on the top an expanded view of the ionization fraction of the elements C, S, Mg, and Fe.

With reference to equation 3, in regions
where H is predominantly neutral, (below about 1500 km altitude)
$\nu _{s,i}\sim\nu _{s,n}$ and similar fractionation
results for elements where $\xi _s$ is reasonably close to unity. Where H is predominantly
ionized, $\nu _{s,i} >> \nu _{s,n}$, and small departures in $\xi _s$ from unity can make
a big difference to the fractionation. This is the reason why S fractionates similarly to
Fe and Mg in the low chromosphere, but markedly less so in higher regions. This is also the
reason why Mg fractionates less than Fe higher up. At an altitude of 2000 km, the charge
state fractions of Fe, Mg, and S are 0.9995, 0.9981, and 0.9942 respectively (see Figure 8).
Even though
these are close to unity, the differences from unity result in different fractionations
where the H ionization fraction (which closely follows that of O) is about 0.6.
Lower down, where H is mostly neutral,
the different ionization fractions matter much less in the fractionation. Recalling the
results calculated using the Saha approximation for the ionization fractions, we can
now see why Mg and Fe fractionate much more similarly in this case. The ionization fractions
at 2000 km altitude are now 0.999974, 0.999995, and 0.9983 respectively, for Fe, Mg, and S.
These are much closer to unity than before, so Fe and Mg now fractionate to a more
similar degree. This is to be expected, since the assumption of LTE in the Saha
equation suppresses radiative recombination rates, since the photons so produced cannot
escape, and so the Saha ionization fractions will be higher than a more realistic
calculation would predict.

However in Table 1 using Saha equilibrium, Fe and Mg do not behave exactly identically.
Recalling equation 3,
\begin{equation}
{\rm fractionation} =\exp\left(2\int
_{z_l}^{z_u}{\xi _sa\nu _{s,n}\over \left[\xi _s\nu _{s,n}+\left(1-\xi _s\right)\nu
_{s,i}\right]}{1\over \left[kT/m_s +v_{\mu turb}^2 +v_{turb}^2\right]}dz\right)
\end{equation}
and remembering that $v_{turb}$ is the amplitude of slow mode waves generated by the
Alfv\'en waves themselves, we can see that when $v_{turb}$ and $v_{\mu turb}$
dominate over the ion thermal speed (usually $v_{turb} > v_{\mu turb}$), the mass dependence disappears from this
part of the equation, and will only reside, if at all, in the collision frequencies.
In fractionation occurring high in the chromosphere associated with the ``spike'', where
the plasma temperature is increasing rapidly up to coronal values,
this condition may not be met and mass dependent fractionation can occur. In fractionation
occurring lower down near the chromospheric temperature minimum, for example in the
0.085 rad s$^{-1}$ case, this condition is met, and quasi-mass independent
fractionation results.

\section{Discussion}
\subsection{The Effect of Slow Mode Waves}
The parametric generation of slow mode wave is a crucial part of the fractionation process
by ponderomotive forces.
One important effect has been to render the fractionation quasi mass independent as is often
observed. This is demonstrated most clearly in Figure 7a, corresponding to a
0.085 rad s$^{-1}$ Alfv\'en wave, the case which also has the highest slow mode wave
amplitude, staying close to 10 km s$^{-1}$ for large regions of the chromosphere where
thermal speeds are $\sim 1$  km s$^{-1}$. Fractionation
occurring at the top of the chromosphere in the location of the ``spike'' in the
ponderomotive acceleration often retains some mass dependence,
because the plasma temperature is
increasing rapidly here while the slow mode wave amplitude is usually small.

In the case that $\delta v_s \sim \delta v_A$ as in \citet{laming09}, the increased
slow  mode wave amplitude relative to this work suppresses all FIP fractionation except
when the wave frequency coincides precisely with the loop resonance, and then all
fractionation occurs at the loop top and hence is mass dependent. For reasons we discuss
below, this coincidence is probably not realized ubiquitously in the solar corona. Moreover
the assumption of isotropic turbulence probably requires a well developed turbulent
cascade, which is unlikely to develop with purely parallel propagating waves
\citep[e.g.][]{luo06}. Lower down in
the chromosphere as the plasma beta approaches unity, the magnetic field becomes
more concentrated in network segments. The increased curvature of field lines will
lead initially parallel propagating waves to develop perpendicular components to their
wavevectors, and hence turbulent cascade or mode conversion become more likely. Our
equation 15 above is an attempt to capture this behavior, and obviously needs to be
revisited with greater rigor.

\subsection{The Alfv\'en Wave Frequency}
Table 5 displays FIP fractionations for a range of frequencies close to the fundamental
of a loop with length 100,000 km and magnetic field 20 G, with wave amplitudes chosen
such that the fractionation of Fe/O is close to the usually observed value of 4. We
commented above how the fractionation details of other elements vary slightly as the
coronal wave moves from being in coincidence with the loop resonance, to a position well
off resonance. This arises because resonant waves reflect from the top of the chromosphere,
and this is then the sole location of FIP fractionation, but nonresonant waves penetrate
further down, allowing FIP fractionation to occur over a greater range of altitudes in
the chromosphere.

When FIP fractionation is concentrated at the top of the chromosphere, the different
ionization structures of the various high FIP elements becomes important, and fractionation
occurs among them. Most significantly, He becomes depleted relative to O, with this
depletion being strongest for a frequency 0.075 rad s$^{-1}$, just higher than the
resonance, at a value of 0.60. This gives an abundance close to the He abundance
frequently observed in the
slow speed solar wind \citep{aellig01,kasper07}. In this frequency region
too, C and possibly S also have minima in their fractionations. These elements have
ionization potentials of 11.26 and 10.36 eV respectively (on
the boundary between low FIP and high FIP elements). Although they are highly ionized
throughout the chromosphere, as described above, they have sufficient neutral component that
they fractionate well when H is predominantly neutral, but not when H is ionized. When
fractionation is restricted to the top of the chromosphere where H is ionized, they behave more
like high FIP elements. This is commonly seen in spectroscopic observations of S
\citep[e.g.][]{laming95,feldman09,widing08,brooks11}.

Off resonance, when FIP fractionation can occur over a more extended range of heights,
including those where H is mainly neutral, C and S might be expected to behave more like
low FIP elements. Such behavior
is more apparent in the solar wind observations of \citet{zurbuchen02} and
\citet{vonsteiger00}. Here, the FIP bias is variable, so that the time average over an
extended period gives Fe/O $\sim 2$ instead of $\sim 4$ as modeled. Even so, S/O has
a similar value to Fe/O. These observations are best matched in Table 5 for a frequency
0.085 rad s$^{-1}$.

We have previously suggested that Alfv\'en waves of coronal origin probably derive from
coronal heating mechanisms such as nanoflares or Alfv\'en resonance. The coronal Alfv\'en
amplitudes required above ($\sim 50 - 100$ km s$^{-1}$) are larger than nonthermal
mass motions observed through spectral linewidths by factors 2-3. This suggests that
the Alfv\'en wave must be confined to a small fraction of the loop cross-sectional area,
which would also be a natural consequence of nanoflare or Alfv\'en resonance heating.

In as far as the
heating can be considered a small perturbation to the coronal loop, the waves so
generated should be eigenfunctions of the loop, with frequencies constrained to coincide
with the loop resonance(s). The fact that many observations are better matched by
Alfv\'en wave frequencies slightly higher than the resonance possibly suggests a
dynamic system. The loop releases waves at its resonance as part of the heating process.
The heat conducts down to the chromosphere, and heated plasma there evaporates back
upwards into the loop \citep[e.g.][]{patsourakos06},
thus increasing its density and reducing the coronal Alfv\'en speed, and hence
also reducing the loop resonance frequency. The waves produced at the original resonance continue to
propagate until damped. So we might naturally expect a mismatch between Alfv\'en wave
frequency and the loop resonance, and also expect this mismatch to become larger in
more strongly heated region of the corona, e.g. active region and flares, as opposed
to quiet sun. Of course this discussion presupposes that the heating is a weak
perturbation to the coronal loop, and so its eigenfunctions are well defined, and can be
excited. In flares and CMEs, this might no longer be true, and the heating mechanism
itself will
determine which waves are produced, irrespective of the loop boundary conditions.

Such ideas will be investigated in greater detail in a separate paper. For the time being,
we restrict ourselves to some simple predictions. The coronal helium abundance should increase
with increasing solar activity, as it appears to do both in solar wind observations
\citep{aellig01,kasper07} and in spectroscopic measurements of the quiet sun
\citep{laming01,laming03} compared to flares \citep{feldman05}. The S and C abundance
should also vary. In the solar wind \citep[e.g.][]{vonsteiger00} it appears to vary
as a low FIP ion, whereas in spectroscopic observations, \citep[e.g.][of quiet and active
regions]{laming95,feldman09} sulfur is observed to behave as a high FIP element.

\subsection{The Upward Flow Speed}
We have suggested conduction driven chromospheric evaporation as the source of the plasma
upflow that populates the corona with fractionated gas. Here we estimate the flow speed
in the chromosphere, and show that it is consistent with limits set by the operation of
the ponderomotive force producing the FIP fractionation. With $d\rho /dt=0$ we write
\begin{equation}
{\partial\over\partial z}\left(\rho v\right)=-{\partial \rho\over\partial t} =
-{\mu gz\over k_{\rm B}T^2}{\partial T\over\partial t}\rho
\end{equation}
where the density $\rho\propto\exp\left(-\mu gz/k_{\rm B}T\right)$ is a gravitationally
stratified solution. The mean molecular mass is $\mu$, and $k_{\rm B}$ is Boltzmann's
constant. Integrating between $z_l$ and $z_u$
\begin{equation}
\rho\left(z_u\right)v\left(z_u\right)-\rho\left(z_l\right)v\left(z_l\right)=
-\int _{z_l}^{z_u}{\mu gz\over k_{\rm B}T^2}{\partial T\over\partial t}\rho dz.
\end{equation}
We choose the upper limit $z_u$ to be where $v\left(z_u\right)=0$ in the corona, and
so
\begin{equation}
v\left(z_l\right)=\int _{z_l}^{z_u}{z\over L_{\rho}}{\rho\left(z\right)\over\rho\left(z_l\right)}{\partial\ln T\over\partial t}dz
=\int _{z_l}^{z_u}{z\over L_{\rho}}\exp\left(-{z\over L_{\rho}}\right){\partial\ln T\over\partial t}dz
\end{equation}
where $L_{\rho}=k_{\rm B}T/\mu g$ is the density scale height. This integral will
be dominated by the integrand near $z=z_l$, so
\begin{equation}
v\left(z_l\right)\simeq L_{\rho}{\partial\ln T\over\partial t}\simeq {2L_{\rho}\over
5nk_{\rm B}T}{\partial\over\partial z}\left(10^{-6}T^{5/2}{\partial T\over\partial z}\right)\sim 10^{17}{T^{1/2}\over n}\left(\Delta T\over L_T\right)^2
\end{equation}
where we have put $2.5 nk_{\rm B}\partial T/\partial t=-\nabla\cdot{\bf Q}$ with
the heat flux ${\bf Q}=-10^{-6}T^{5/2}\nabla T$. The temperature gradient has been replaced
by $\Delta T/L_T$,
where $\Delta T$ may be the coronal peak temperature, and $L_T$ the loop half-length.
Taking the chromospheric temperature $T\sim 10^4$,
\begin{equation}
v\left(z_l\right)\simeq {10^{13}\over n}\left(T_c/5\times 10^6~{\rm K}\over L_T/5\times 10^9~{\rm cm}\right)^2 {\rm cm s}^{-1},
\end{equation}
which suggests a velocity of $10^3$ cm s$^{-1}$ at a density of $10^{10}$ cm$^{-3}$.
This may well be an underestimate due to our approximation for the temperature gradient,
but as discussed in \citet{laming04}, is sufficiently high that gravitational settling
should not occur. If $v\left(z_l\right)$ approaches $\sim 1$ km s$^{-1}$, as might happen
in flares, some further discussion is required.

The derivation of equation 17 neglected inertial terms in the momentum equations for
ions and neutrals. Reinstating these, in the limit that $u_{si}-u_{sn} << u_s\sim u$ the
fractionation becomes
\begin{equation}
{\rm fractionation} =\exp\left(2\int
_{z_l}^{z_u}{\xi _sa\nu _{s,n}\over \left[\xi _s\nu _{s,n}+\left(1-\xi _s\right)\nu
_{s,i}\right]}{1\over \left[kT/m_s +v_{\mu turb}^2 +v_{turb}^2+u_{flow}^2\right]}dz\right).
\end{equation}
When the magnitude of $u_{flow}^2$ approaches those of the other terms in the second
square bracket in the denominator of the integrand, some reduction in the FIP effect
will result. This is most likely to have an impact on fractionation occurring at the
top of the chromosphere, taking $\rho u_{flow}\sim $ constant through the chromosphere,
and will possibly reduce
the amount of mass dependent fractionation occurring there. This will happen for
relatively large flow speeds $u_{flow}\sim 1$ km s$^{-1}$ or larger at the top of the
chromosphere, still significantly lower than the Alfv\'en speed in this region.

\subsection{Conclusions}
We have further developed the model of element fractionation to give rise to the FIP effect
by the ponderomotive force, paying careful attention to the generation of slow mode waves
by the primary Alfv\'en oscillations. When considering a realistic wave spectrum with both
chromospheric and and coronal contributions, the extra longitudinal pressure due to the
slow mode waves is crucial in producing the correct fractionation. With this extra
ingredient, together with the improvements to the ionization balance and the normalization
of the fractionation, a rather comprehensive description of the coronal fractionation
has emerged. In seeking to understand the FIP effect as usually described, we have also
found an explanation for the depletion of He in the solar wind, and also possibly its
variation. The Ne abundance also appears to vary in a similar manner, but to a lesser degree.
It is also more sensitive to assumptions about the ionization balance, but further
investigation of this is expected to resolve current controversy surrounding the solar
photospheric abundance of this element.

The theory now appears to be developed to the point where variations in the FIP fractionation
from place to place in the solar corona or wind may now be interpreted in terms of their
physical origins. The element abundances in the solar corona may therefore be considered
as diagnostics of the behavior of MHD turbulence, and also thereby of the mechanisms that
heat the solar corona. These ideas will be further developed in subsequent papers.

\acknowledgements
This work was supported by NASA Contracts NNH10A055I, NNH11AQ23I, and by basic research funds of
the Office of Naval Research. I am also grateful to Cara Rakowski for a critical reading
of an earlier draft of this paper.

\clearpage
\begin{table}[t]
\begin{center}
\caption{FIP Fractionations in Different Approximations}
%\scriptsize
\begin{tabular}{lrrrr}
\tableline\tableline
ratio& baseline$^a$ & density mod.$^b$ & Saha ionization$^c$& $\delta v_z=\delta V_A$\\ \tableline
He/O & 0.67& 0.83& 0.73& 0.85\\
C/O & 0.99& 1.26& 1.17& 1.03\\
N/O & 0.82& 1.02& 0.95& 0.93\\
Ne/O& 0.74& 0.93& 0.90& 0.89\\
Mg/O& 1.98& 2.52& 2.33& 1.43\\
Si/O& 1.89& 2.37& 2.41& 1.41\\
S/O & 1.40& 1.75& 1.47& 1.23\\
Ar/O& 0.92& 1.16& 1.03& 0.97\\
Fe/O& 3.29& 4.17& 2.79& 1.87\\
\tableline
\end{tabular}
\end{center}
\tablecomments{$^a$ Baseline model corresponds to a coronal peak Alfv\'en wave
amplitude of $\sim 60$ km s$^{-1}$, and the standard chromospheric model. $^b$ gives results with chromosphere density modified to be consistent with
photoionization-recombination equilibrium and the temperature in the \citet{avrett08} model. $^c$ give results with the ionization fraction for each
element modified to be given by Saha equilibrium at the temperature and density
given by \citet{avrett08}. The final column shows the effect of the approximation of \citet{laming09}, where isotropic turbulence is assumed.}
\label{tab1}
\end{table}

\begin{table}[t]
\begin{center}
\caption{Open Magnetic Field Wave Spectra at 500,000 km Altitude}
%\scriptsize
\begin{tabular}{lrrr}
\tableline\tableline
ang. freq. & v0 & v1& v2 \\ \tableline
0.010 & 12.5& 12.5& 125\\
0.031 & 150& 150& 150\\
0.062 & 75& 75& 75\\
0.093&  50& 50& 50\\
0.124&  12.5& 125& 12.5\\
\tableline
\end{tabular}
\end{center}
\tablecomments{Frequencies are angular frequencies in rad s$^{-1}$. Velocities
are in km s$^{-1}$; v0 is designed to match Fig. 3 in \citet{cranmer07}.}
\label{tab2}
\end{table}

\begin{table}[t]
\begin{center}
\caption{FIP Fractionations in Open Magnetic Field}
%\scriptsize
\begin{tabular}{l|rrr|rrr}
\tableline\tableline &
\multicolumn{3}{c}{models}&\multicolumn{3}{c}{observations}\\
ratio & v0& v1& v2& a& b& c\\ \tableline
He/O & 0.90& 0.85& 0.85&  0.60-0.58& 0.37-0.47& 0.45-0.55\\
C/O & 1.13& 1.18& 1.10&   1.50-1.41& 1.17-1.35& 0.9 - 1.1\\
N/O & 0.96& 0.94& 0.94&    1.19-0.9& 0.64-0.99\\
Ne/O& 0.95& 0.92& 0.92&   0.48-0.47& 0.40-0.56& 0.3 - 0.4\\
Na/O& 1.97& 2.99& 2.04&    \\
Mg/O& 1.65& 2.21& 1.67&   1.73-1.92& 0.98-1.60& 0.95 - 2.45\\
Al/O& 1.72& 2.37& 1.75&      \\
Si/O& 1.51& 1.89& 1.53&    2.07-1.92& 1.20-2.09& 0.9 - 1.8\\
S/O & 1.26& 1.39& 1.26&    1.53-1.56& 1.38-2.57\\
Ar/O& 1.00& 0.99& 0.99&    \\
K/O & 2.03& 3.14& 2.12&    \\
Ca/O& 1.88& 2.74& 1.93&  \\
Fe/O& 1.97& 2.97& 2.04&  1.42-1.73& 1.04-1.69 &0.65 - 1.35\\
Ni/O& 1.85& 2.67& 1.91&   \\
Kr/O& 1.01& 1.01& 1.01&   \\
Rb/O& 1.97& 2.94& 2.04&    \\
W/O & 1.99& 2.99& 2.07&  \\
\tableline
\end{tabular}
\end{center}
\tablecomments{From left to right, model FIP fractionations correspond
  to the chromospheric model in Fig. 3.  Observational ratios are
  taken from, (a) Zurbuchen
  et al. (2002), (b) von Steiger et al. (2000), (c) Ko et al. (2006), all given relative to O.
  Ranges quoted from von Steiger et al. (2000) include their uncertainties.}
\label{tab3}
\end{table}

\begin{table}[t]
\begin{center}
\caption{FIP Fractionations in Closed Magnetic Field I}
%\scriptsize
\begin{tabular}{l|rrr|rrrrrr}
\tableline\tableline &
\multicolumn{3}{c}{models}&\multicolumn{6}{c}{observations} \\
& 0.4& 0.5& 0.6& a& b& c& d& e& f\\
ratio & \multicolumn{3}{c}{(km s$^{-1}$)}& \\\tableline
He/O & 0.72& 0.61& 0.47&  0.68-0.60& 0.29-0.75& & & & \\
C/O & 1.01& 1.03& 0.97&  1.36-1.41& 1.06-1.37& & & & \\
N/O & 0.86& 0.81& 0.69&  0.72-1.32& 0.22-0.89& & & & \\
Ne/O& 0.86& 0.71& 0.57&  0.58& 0.38-0.75& & & & \\
Na/O& 2.42& 3.96 & 6.74&  & & 7.8${+13\atop -5}$&
1.8${+2\atop -1}$\\
Mg/O& 1.76& 2.35&  3.25& 2.58-2.61& 1.08-2.36& 2.8${+2.3\atop -1.3}$& 2.7$\pm 0.3$& &\\
Al/O& 1.95& 2.82&  4.12& & & 3.6${+1.7\atop -1.2}$&
5.6${+3.3\atop -2.1}$\\
Si/O& 1.70& 2.27&  3.02& 2.49-3.11& 1.36-3.24& 4.9${+2.9\atop -1.8}$& & & \\
S/O & 1.34& 1.57&  1.78& 1.62-1.92& 1.23-2.68& 2.2$\pm 0.2$& 2.1$\pm 0.2$& $1.7\pm
0.3$& \\
Ar/O& 0.96& 0.94&  0.86& & & & & $1.1\pm 0.1$& 1.12$\pm 0.15$\\
K/O & 2.72& 4.70& 8.51&  & &  1.8${+0.4\atop -0.6}$&
4.7${+7.0\atop -2.8}$&  $3.5\pm 0.9$& 6\\
Ca/O& 2.38& 3.80&  6.27& & & 3.5${+4.3\atop -1.9}$&
2.7$\pm 0.25$& & 3.0-9.7\\
Fe/O& 2.65& 4.44&  7.85& 2.28-2.90& 0.96-2.46& 7.0${+1.4\atop -1.2}$& & & \\
Ni/O& 2.59& 4.10&  6.92& & & & & \\
Kr/O& 0.99& 1.00&  0.92& & & & & \\
Rb/O& 2.84& 4.96&  9.01 & & & & & \\
W/O & 3.01& 5.35&  9.85& & & & & \\
\tableline
\end{tabular}
\end{center}
\tablecomments{From left to right, model FIP fractionations correspond
to the chromospheric model in Fig. 3. The models for 0.4, 0.5 and 0.6 km s$^{-1}$ refer to the
amplitude with which the coronal wave is initiated in the low chromosphere. The corresponding
coronal wave amplitudes are 55, 67, and 82 km s$^{-1}$ respectively.
Observational ratios are taken from, (a) Zurbuchen
et al. (2002), given relative to O, (b) von Steiger et al. (2000), relative to O, (c) Bryans et al. (2009), given
relative to the mean of O, Ne and Ar, (d) Giammanco et al. (2007, 2008),
relative to H, and (e) Phillips et al. (2003), relative to H, (f)
Sylwester et al. (2010ab) and McKenzie \& Feldman (1994), relative to H, all given relative to the
photospheric abundance of Grevesse \& Sauval (1998). Ranges quoted from von Steiger et al. (2000) include their uncertainties.}
\label{tab4}
\end{table}

\begin{table}[t]
\begin{center}
\caption{FIP Fractionations in Closed Magnetic Field II}
%\scriptsize
\begin{tabular}{|l|rrrrrrrrr|}
\tableline\tableline ang. freq. (rad s$^{-1}$) & 0.055& 0.06& 0.065& 0.07& 0.075& 0.08& 0.085&0.09&0.105\\
$v_{init}$ (km s$^{-1}$)& 0.255& 0.25& 0.35& 0.45& 0.55&0.5&0.45&0.36&0.193\\
$v_{cor}$ (km s$^{-1}$)& 40 & 40& 45& 60& 70& 60& 50& 38& 20\\
ratio & & & & & & & & &\\ \tableline
He/O& 0.84& 0.81& 0.74& 0.63& 0.60& 0.74& 0.86& 0.94& 0.98\\
C/O & 1.43& 1.38& 1.32& 1.18& 1.23& 1.40& 1.53& 1.59& 1.56\\
N/O & 0.93& 0.92& 0.89& 0.83& 0.80& 0.88& 0.94& 0.97& 0.99\\
Ne/O& 0.90& 0.89& 0.84& 0.75& 0.72& 0.83& 0.91& 0.96& 0.98\\
Na/O& 4.74& 4.47& 4.61& 4.31& 4.47& 4.36& 4.69& 4.73& 4.60\\
Mg/O& 3.29& 3.13& 3.11& 2.74& 2.89& 3.13& 3.49& 3.59& 3.51\\
Al/O& 3.59& 3.43& 3.49& 3.15& 3.30& 3.42& 3.73& 3.81& 3.75\\
Si/O& 2.70& 2.63& 2.69& 2.51& 2.69& 2.74& 2.88& 2.89& 2.84\\
S/O & 1.79& 1.77& 1.79& 1.72& 1.83& 1.86& 1.92& 1.90& 1.87\\
Ar/O& 0.98& 0.98& 0.97& 0.96& 0.94& 0.97& 0.99& 0.99& 1.00\\
K/O & 5.08& 4.85& 5.13& 4.84& 4.92& 4.68& 4.95& 4.98& 4.94\\
Ca/O& 4.31& 4.13& 4.32& 4.00& 4.12& 4.06& 4.34& 4.39& 4.36\\
Fe/O& 4.78& 4.60& 4.87& 4.52& 4.57& 4.44& 4.73& 4.79& 4.79\\
Ni/O& 4.20& 4.09& 4.35& 4.12& 4.23& 4.06& 4.24& 4.25& 4.27\\
Kr/O& 1.00& 1.00& 1.00& 1.00& 0.99& 1.00& 1.00& 1.00& 1.00\\
Rb/O& 4.73& 4.61& 4.98& 4.79& 4.80& 4.50& 4.67& 4.67& 4.72\\
W/O & 4.84& 4.73& 5.15& 4.83& 4.83& 4.56& 4.76& 4.79& 4.89\\
\tableline
\end{tabular}
\end{center}
\tablecomments{Entries in rows $v_{init}$ and $v_{cor}$ refer to the
amplitude with which the coronal wave is initiated in the low chromosphere, and it corresponding
amplitude in the corona.}
\label{tab5}
\end{table}

\clearpage

\begin{figure}
\epsscale{0.75} \plotone{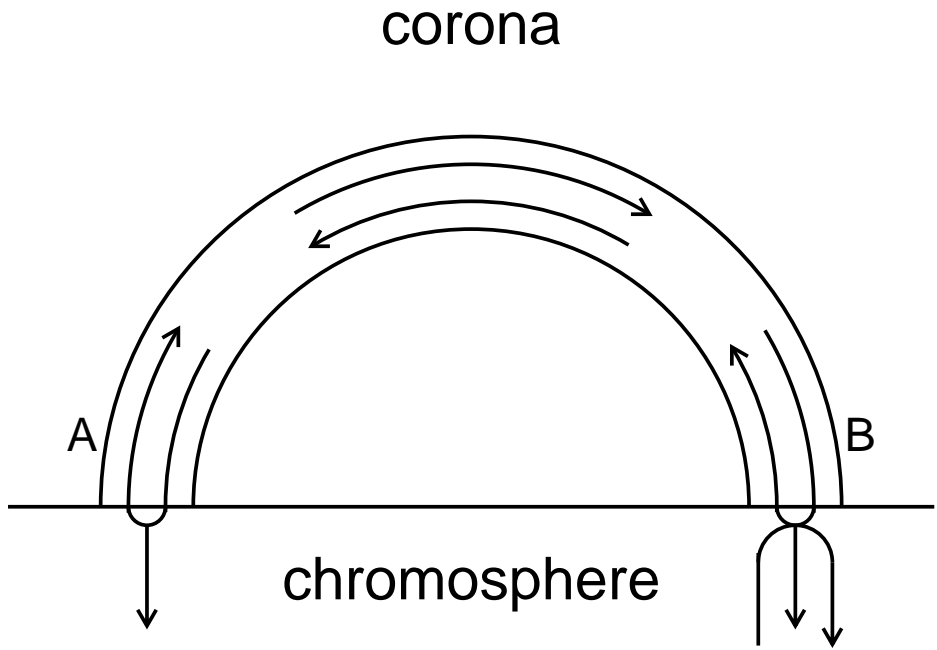} \figcaption[f1.eps]{Cartoon
illustrating the model. Alfv\'en waves generated in the corona either reflect
from each footpoint or precipitate down, depending on their frequency with respect to the loop resonance. Resonant waves reflect, nonresonant waves precipitate down.
Upcoming waves in the chromosphere, deriving from the mode or parametric conversion of p-modes at the $\beta =1.2$ layer, are generally reflected back downwards, as
illustrated at footpoint B. In our models, we specify wave amplitudes at footpoint A, and integrate the non-WKB transport equation back to footpoint B, where FIP fractionations are evaluated.
\label{fig1}}
\end{figure}

\begin{figure}[t]
\epsscale{1.0} \plotone{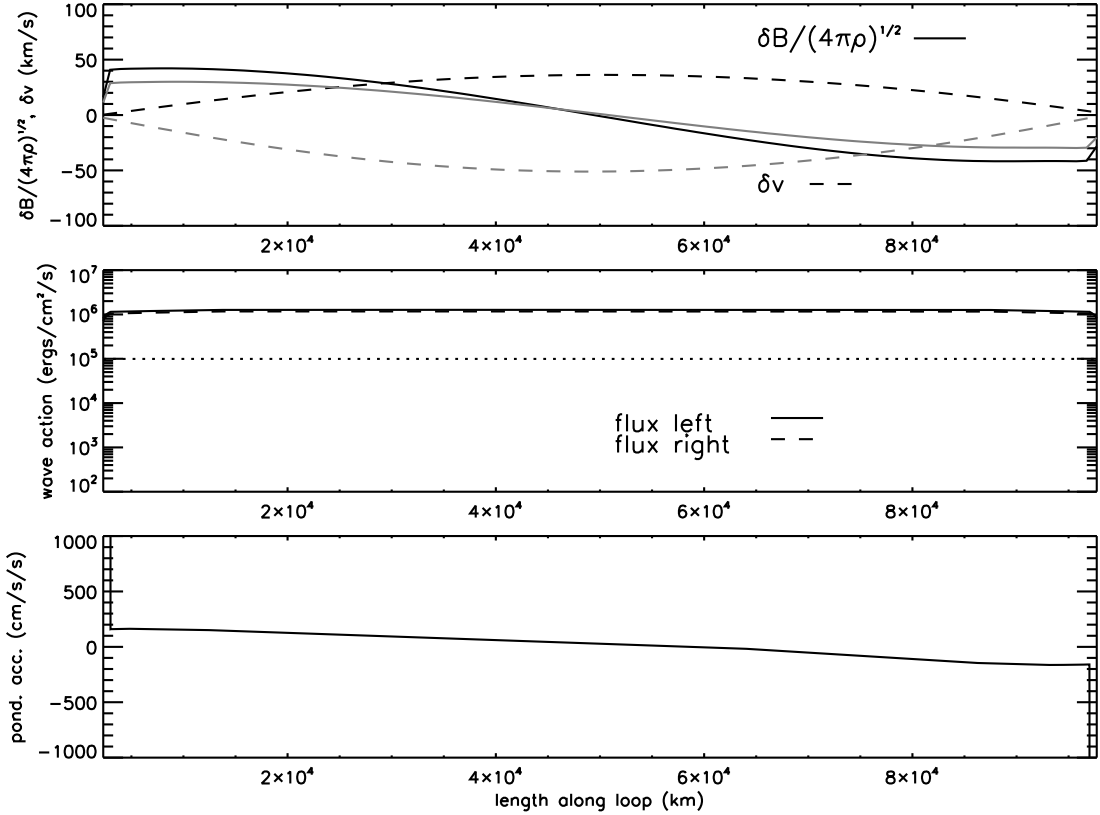} \figcaption[f2.ps]{Coronal section of
loop, length 100,000 km, magnetic field 20 G, (half wavelength
long for a 0.07 rad s$^{-1}$ angular frequency Alfv\'en wave) showing from top: Els\"asser variables in km s$^{-1}$ ($\delta
B/\sqrt{4\pi\rho}$ solid lines, $\delta v$ dashed lines), with black
lines for real parts and gray lines for imaginary parts. Middle; wave energy fluxes in ergs cm$^{-2}$ s$^{-1}$, the thin solid line shows the difference in
energy fluxes divided by the magnetic field strength and should be a
horizontal line if energy is properly conserved. Bottom, the
ponderomotive acceleration in cm s$^{-2}$. Positive acceleration means
positive along the $z$ axis, which is upwards pointing near $z=0$ and
downwards near $z=100,000$.\label{fig2}}
\end{figure}

\begin{figure}[t]
\epsscale{1.0} \plotone{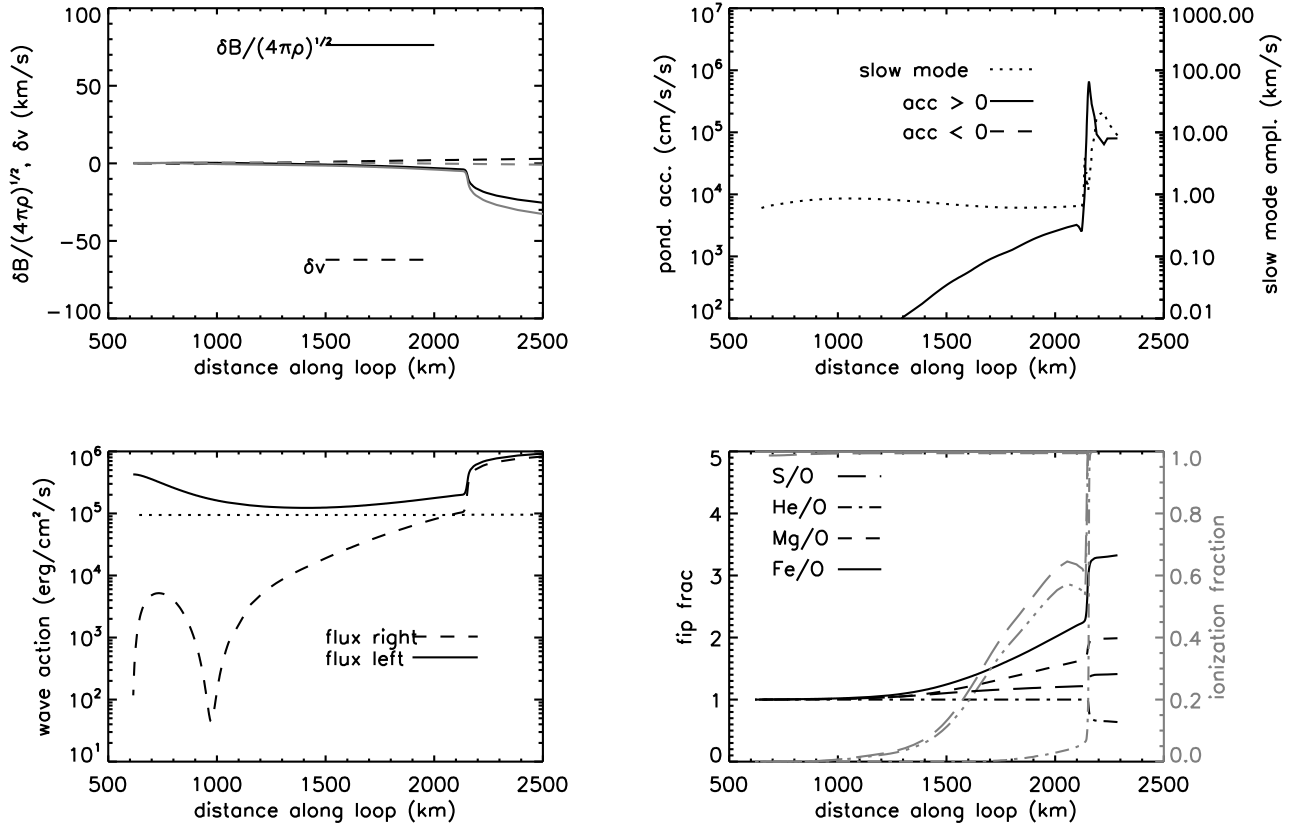} \figcaption[f3.ps]{Same as figure 2
giving the first three panels for the left hand chromosphere ``B'', where
waves leak down from the corona. The extra bottom right panel shows
the FIP fractionations (in black) for the ratios Fe/O, Mg/O, S/Oand He/O.
Chromospheric ionization fractions are also shown in the fourth panel
(in gray, to be read on the left hand axis) in the same linestyles as for the fractionation.
The extra dash-triple dot gives the O ionization fraction, the long dash curve gives the
H ionization fraction. Fe and Mg are essentially fully ionized throughout the fractionation region.\label{fig3}}
\end{figure}

\begin{figure}[t]
\epsscale{1.0} \plotone{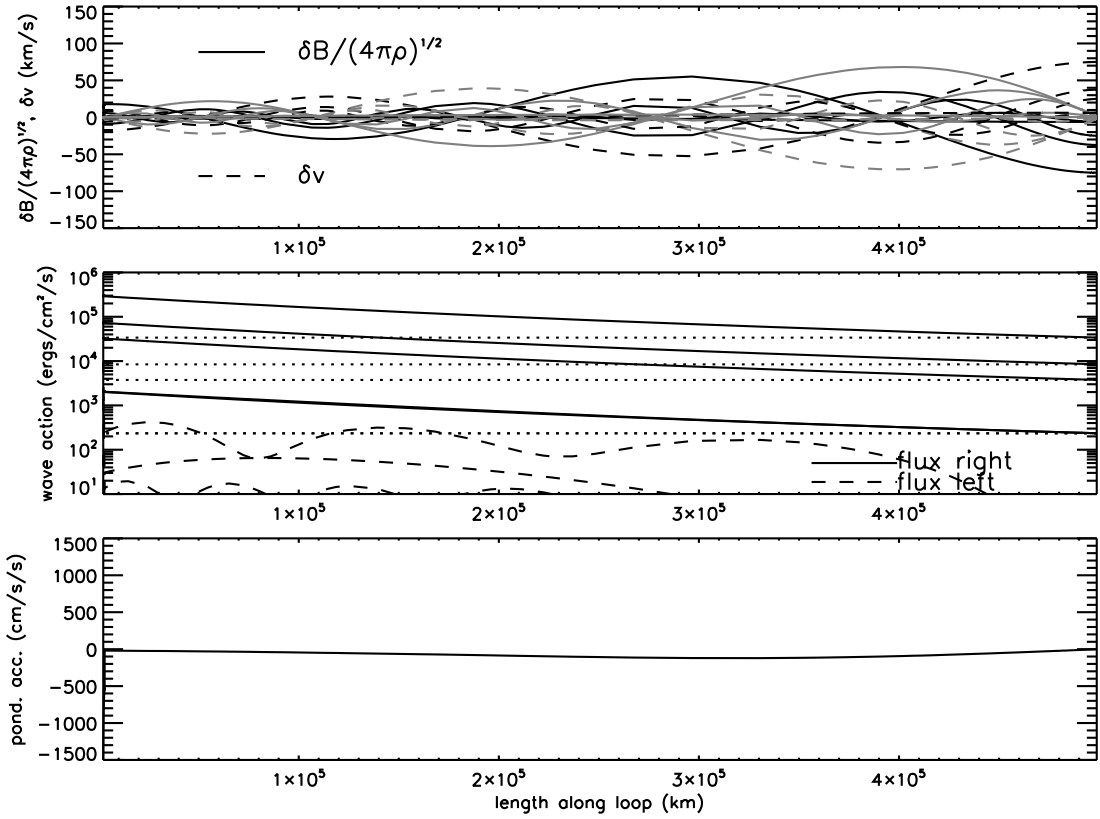} \figcaption[f4.ps]{Same as figure 2
for an open field region. Five waves corresponding to the baseline model in Table 2 are considered, initiated at 500,000 km altitude and integrated back to the chromosphere.\label{fig4}}
\end{figure}

\begin{figure}[t]
\epsscale{1.0} \plotone{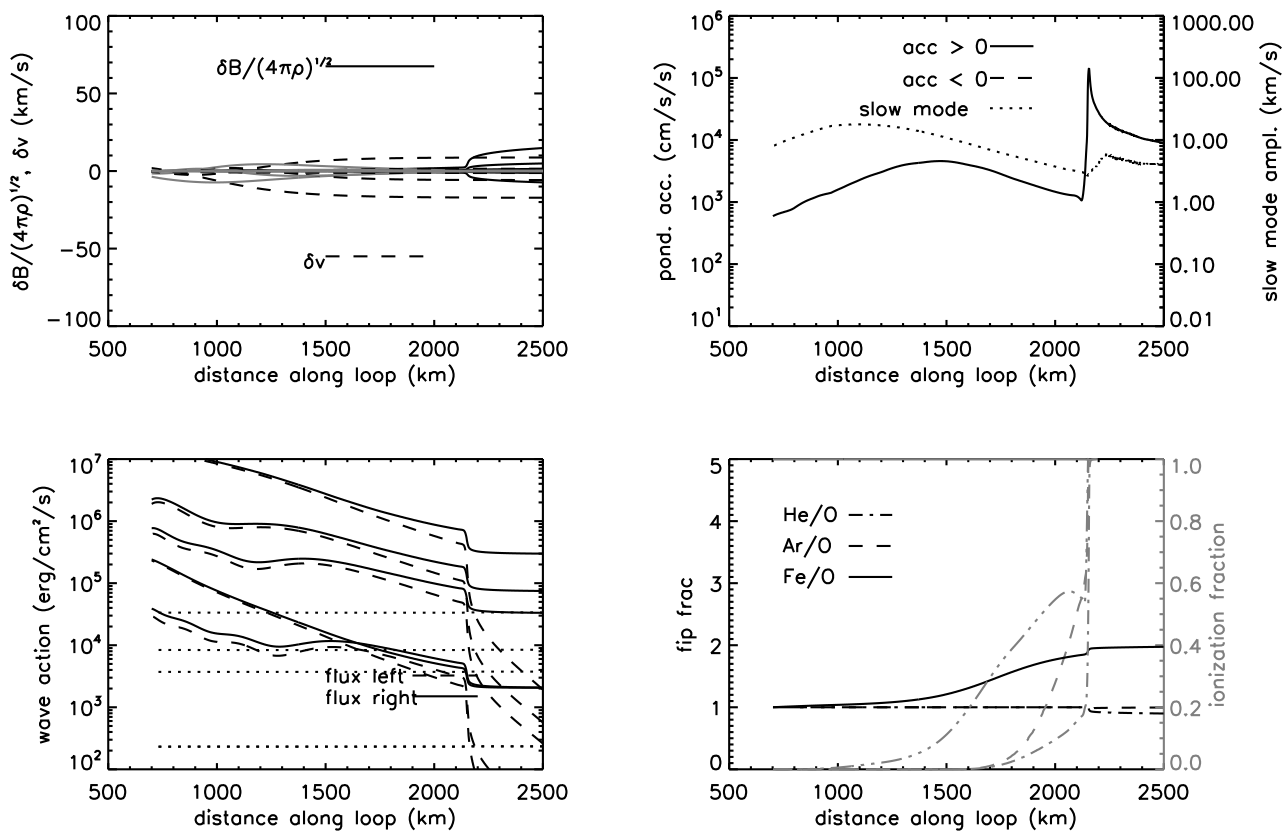} \figcaption[f5.ps]{Chromospheric section of open field calculation of Figure 4. Fractionations are shown for Fe/O, Ar/O and He/O.
\label{fig5}}
\end{figure}

\begin{figure}[t]
\epsscale{1.0}
\includegraphics[width=6in,angle=-90]{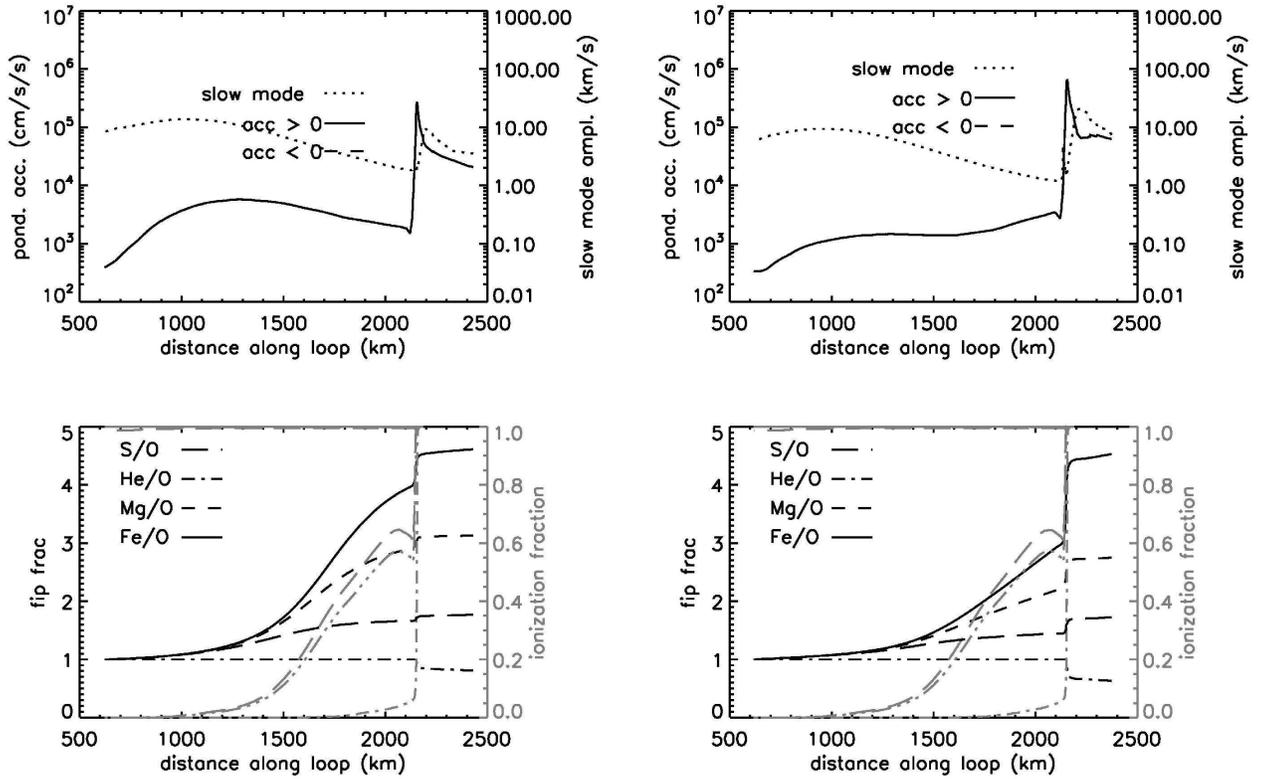} \figcaption[f6.eps]{\label{fig6}Ponderomotive
force (top panels) and FIP fractionations (bottom panels) for the chromosphere including five chromospheric waves and a coronal wave with angular frequency
0.06 (left) and 0.07 rad s$^{-1}$ (right) respectively. Fractionations for
Fe/O, Mg/O, S/O, and He/O are shown.}
\end{figure}

\begin{figure}[t]
\epsscale{1.0}
\includegraphics[width=6in,angle=-90]{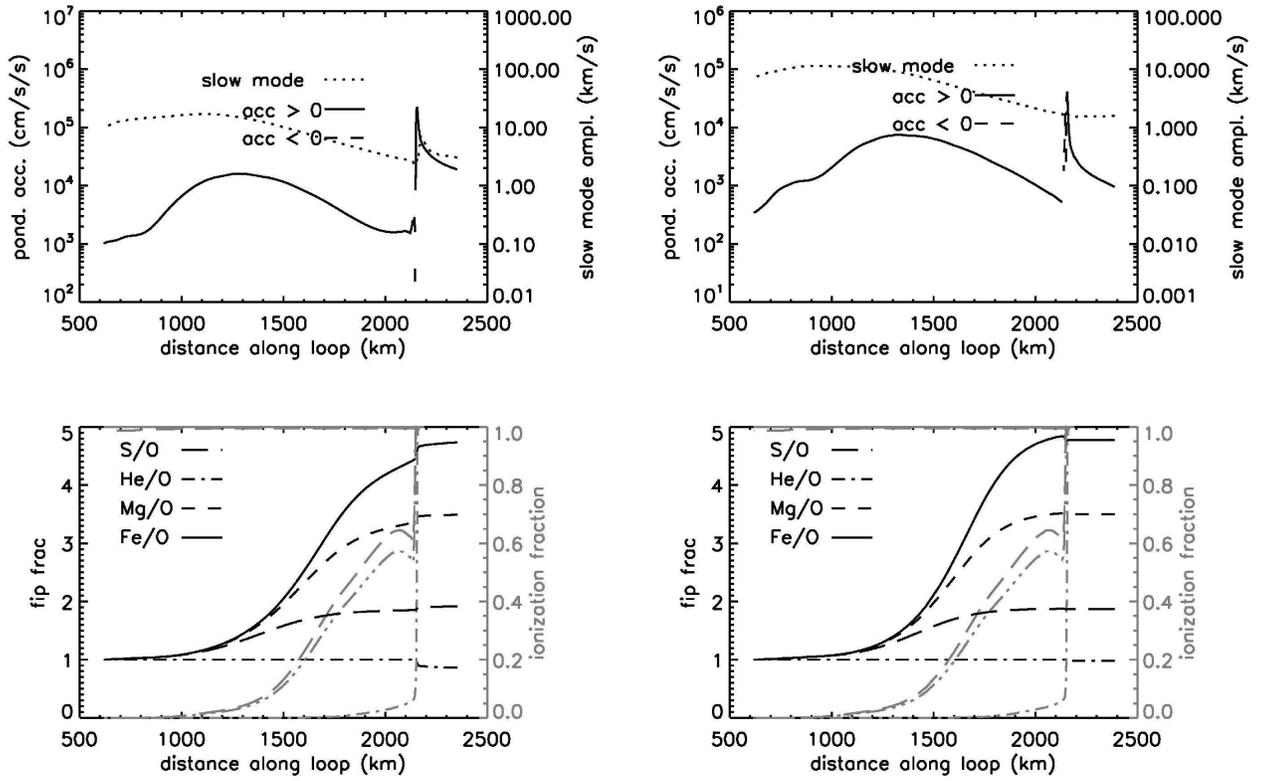} \figcaption[f7.eps]{\label{fig7}Ponderomotive
force (top panels) and FIP fractionations (bottom panels) for the chromosphere including five chromospheric waves and a coronal wave with angular frequency
0.085 (left) and 0.0105 rad s$^{-1}$ (right) respectively. Fractionations for
Fe/O, Mg/O, S/O, and He/O are shown. The 0.085 coronal wave gives a more similar FIP
effect for Fe/O and Mg/O, as frequently observed. The He/O depletion reduces
as the coronal wave moves off resonance, as the ``spike'' in the ponderomotive
acceleration decreases in prominence.}
\end{figure}

\begin{figure}[t]
\epsscale{1.0} \plotone{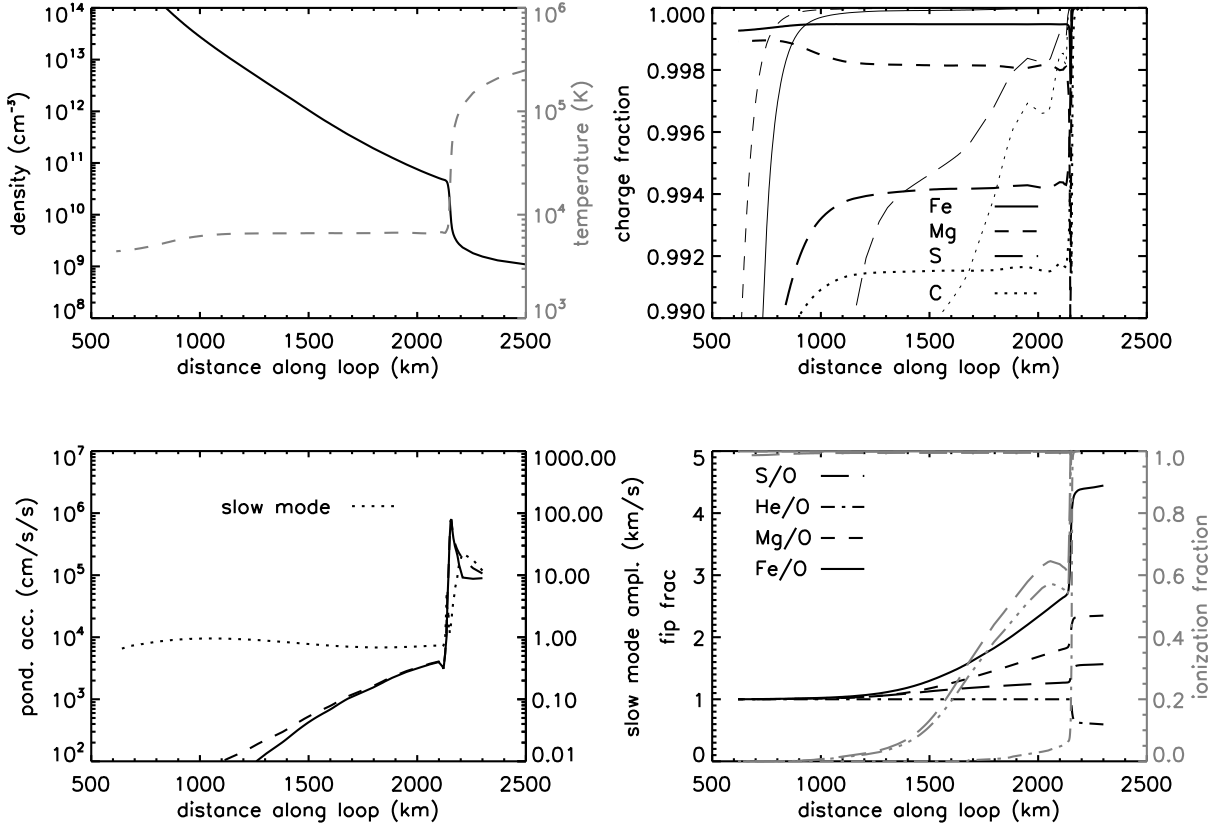} \figcaption[f8.ps]
{Illustration of FIP fractionation for the 0.07 rad s$^{-1}$ wave, showing the correspondence
with important features of the chromosphere. Top left gives the chromospheric density and
temperature with height. Bottom left gives the ponderomotive acceleration as before, with solid
and dashed curves showing the acceleration with and without energy loss to slow mode waves. The 
dotted curve shows the slow mode wave amplitude.
The left panels show the ``spike'' in the ponderomotive
acceleration at the altitude where the chromospheric density gradient is strongest.
Bottom right is the same as before, with FIP fractionations for Fe/O, Mg/O, He/O, and S/O, together
with ionization fractions. Top right shows the ionization fractions for C, S, Mg, and Fe in an
expanded view. Thick curves correspond to the ``baseline'' charge state fractions used for
FIP fractionation throughout this paper, thin curves give the results of the Saha approximation.
This overestimates ionization fraction close to the top of the chromosphere, because photons
resulting from radiative recombinations are not allowed to escape, but remain trapped to cause
further photoionization.
\label{fig8}}
\end{figure}

\end{document}